\documentclass[preprint,3p,twocolumn]{elsarticle}

\usepackage[T1]{fontenc}
\usepackage{amsmath}
\usepackage{newtxtext}
\usepackage{newtxmath}
\usepackage{microtype}
\usepackage{booktabs}
\usepackage{xcolor}
\usepackage[
    colorlinks=true,
    urlcolor=blue,
    linkcolor=black,
    citecolor=black
]{hyperref}

\journal{Nuclear Physics B}

\begin{document}

\begin{frontmatter}

%% Title, authors and addresses

%% use the tnoteref command within \title for footnotes;
%% use the tnotetext command for theassociated footnote;
%% use the fnref command within \author or \affiliation for footnotes;
%% use the fntext command for theassociated footnote;
%% use the corref command within \author for corresponding author footnotes;
%% use the cortext command for theassociated footnote;
%% use the ead command for the email address,
%% and the form \ead[url] for the home page:
%% \title{Title\tnoteref{label1}}
%% \tnotetext[label1]{}
%% \author{Name\corref{cor1}\fnref{label2}}
%% \ead{email address}
%% \ead[url]{home page}
%% \fntext[label2]{}
%% \cortext[cor1]{}
%% \affiliation{organization={},
%%             addressline={},
%%             city={},
%%             postcode={},
%%             state={},
%%             country={}}
%% \fntext[label3]{}

\title{\textbf{iBitter-Stack: A Multi-Representation Ensemble Learning Model\\ for Accurate Bitter Peptide Identification}}

%% use optional labels to link authors explicitly to addresses:
%% \author[label1,label2]{}
%% \affiliation[label1]{organization={},
%%             addressline={},
%%             city={},
%%             postcode={},
%%             state={},
%%             country={}}
%%
%% \affiliation[label2]{organization={},
%%             addressline={},
%%             city={},
%%             postcode={},
%%             state={},
%%             country={}}

\author[1]{Sarfraz Ahmad\corref{cor1}}
\cortext[cor1]{Corresponding Author}
\ead{sarfaraz\_076@outlook.com}
\author[1]{Momina Ahsan}
\author[2]{Muhammad Nabeel Asim}
\author[2]{\\Andreas Dengel}
\author[1]{Muhammad Imran Malik}

%% Author affiliation
\affiliation[1]{organization={National University of Sciences and Technology (NUST)},%Department and Organization
            addressline={H-12}, 
            city={Islamabad},
            country={Pakistan}}

%% Author affiliation
\affiliation[2]{organization={German Research Center for Artificial Intelligence (DFKI)},%Department and Organization
            city={Kaiserslautern, 67663},
            country={Germany}} 

%% Abstract
\begin{abstract}

The identification of bitter peptides is crucial in various domains, including food science, drug discovery, and biochemical research. These peptides not only contribute to the undesirable taste of hydrolyzed proteins but also play key roles in physiological and pharmacological processes. However, experimental methods for identifying bitter peptides are time-consuming and expensive. With the rapid expansion of peptide sequence databases in the post-genomic era, the demand for efficient computational approaches to distinguish bitter from non-bitter peptides has become increasingly significant. In this study, we propose a novel stacking-based ensemble learning framework aimed at enhancing the accuracy and reliability of bitter peptide classification. Our method integrates diverse sequence-based feature representations and leverages a broad set of machine learning classifiers. The first stacking layer comprises multiple base classifiers, each trained on distinct feature encoding schemes, while the second layer employs logistic regression to refine predictions using an eight-dimensional probability vector. Extensive evaluations on a carefully curated dataset demonstrate that our model significantly outperforms existing predictive methods, providing a robust and reliable computational tool for bitter peptide identification. Our approach achieves an accuracy of 96.09\% and a Matthews Correlation Coefficient (MCC) of 0.9220 on the independent test set, underscoring its effectiveness and generalizability. To facilitate real-time usage and broader accessibility, we have also developed a user-friendly web server based on the proposed method, which is freely accessible at \href{https://ibitter-stack-webserver.streamlit.app/}{\texttt{ibitter-stack-webserver.streamlit.app}}. This tool enables researchers and practitioners to conveniently screen peptide sequences for bitterness.

\end{abstract}

%% Keywords
\begin{keyword}

bitter peptides \sep bioinformatics \sep sequence classification \sep machine learning \sep protein language models \sep logistic regression 

\end{keyword}

\end{frontmatter}

\section{Introduction}
\label{sec:introduction}

Bitter taste perception serves as a vital biological defense mechanism, allowing humans and other animals to detect and avoid potentially harmful substances, such as environmental toxins and poisonous plants~\cite{drewnowski2000bitter}. Nevertheless, numerous naturally occurring bitter compounds, including alkaloids, polyphenols, and peptides, hold significant value in nutrition and medicine~\cite{matoba1972relationship}. The presence of hydrophobic amino acids, particularly at the C-terminal region, has been strongly associated with peptide bitterness~\cite{matoba1972relationship}. Bitter peptides, typically produced during protein hydrolysis, are especially important in food science and pharmaceutical development. 

However, their identification remains a complex and challenging task. In light of the growing interest in functional foods and bioactive compounds, there is a pressing need to develop rapid and accurate methods for the identification of bitter peptides. Traditional experimental techniques for characterizing bitter peptides, such as biochemical assays, human sensory evaluation, and chromatography-based separation, have been widely utilized~\cite{van2002ftir}~\cite{kim2006application}. Although effective, these methods are labor intensive, time-consuming, and costly~\cite{karametsi2014identification}~\cite{liu2014identification}~\cite{asim2020k}~\cite{asim2022circ}~\cite{nabeel2023dna}~\cite{asim2025peptide}. Furthermore, human sensory testing introduces subjectivity and inter-individual variability, which hampers reproducibility. To address these challenges, computational approaches, particularly machine learning (ML)-based methods, have emerged as powerful alternatives for predicting bitter peptides based on sequence and structural properties~\cite{le2021radiomics}~\cite{ramzan2021machine}~\cite{asim2025peptide}.

Among computational strategies, quantitative structure–activity relationship (QSAR) modeling has been widely used to predict peptide bitterness~\cite{kim2006quantitative}~\cite{wu2007quantitative}~\cite{yin2010studying}~\cite{lin2008new}~\cite{tong2008novel}~\cite{liang2009using}. QSAR models use ML algorithms such as Support Vector Machines (SVM), Artificial Neural Networks (ANN), and Multiple Linear Regression (MLR) to establish mathematical relationships between peptide descriptors and their biological activity~\cite{wu2007quantitative}~\cite{pripp2007modelling}. For example, Yin et al.~\cite{yin2010studying} developed 28 QSAR models using support vector regression (SVR) to estimate peptide bitterness, while Soltani et al.~\cite{soltani2013qsbr} analyzed bitterness thresholds for over 229 peptides using three different ML methods. Open-access tools like \textit{BitterX}~\cite{huang2016bitterx} and \textit{BitterPredict}~\cite{dagan2017bitter} further advanced the field by employing ML classification techniques to identify bitter compounds with high accuracy. These developments underscore the growing utility of data-driven models in peptide bitterness prediction.

Despite these developments, sequence-based ML predictors for bitter peptides remain limited. One of the earliest sequence-based predictors, \textit{iBitter-SCM}~\cite{charoenkwan2020ibitter}, utilized dipeptide propensity scores for bitterness prediction, achieving high accuracy in independent validation tests. However, reliance on single-type feature representations, such as Di-Peptide Composition (DPC), restricted its generalizability~\cite{charoenkwan2020ibitter}. Later, Deep Learning (DL)-based models such as \textit{BERT4Bitter}~\cite{charoenkwan2021bert4bitter} applied Natural Language Processing (NLP) techniques to extract feature representations directly from raw peptide sequences. Although this approach improved prediction accuracy, it lacked integration with physicochemical properties, which are critical for understanding the underlying biochemical mechanisms of bitterness~\cite{manavalan2019mahtpred}~\cite{manavalan2019meta}~\cite{charoenkwan2013hcs}~\cite{charoenkwan2019iqsp}~\cite{liu2020imrm}~\cite{liu2020irna5hmc}.

To overcome these limitations, several recent frameworks have explored multi-representation learning. For instance, \textit{iBitter-Fuse}~\cite{charoenkwan2021ibitter} introduced an ML pipeline that integrates multiple feature encoding schemes, including Dipeptide Composition (DPC), Amino Acid Composition (AAC), Pseudo Amino Acid Composition (PAAC), and physicochemical properties. A genetic algorithm with a self-assessment-report (GA-SAR) was employed for feature selection, leading to an SVM-based classifier that delivered superior performance. Comparative evaluations demonstrated that iBitter-Fuse significantly outperformed traditional classifiers, achieving an independent test accuracy of 0.930, sensitivity of 0.938, specificity of 0.922, and a Matthews Correlation Coefficient (MCC) of 0.859. However, the relatively lower MCC compared to more recent approaches highlights a potential limitation in relying solely on handcrafted features; incorporating NLP-based pre-trained embeddings alongside physicochemical and composition-based strategies could have further improved the model’s robustness and generalizability.

\begin{table*}[!ht]
\caption{Performance of Existing Bitter Peptide Prediction Methods on Independent Test Sets}
\label{tab:literature}
\centering
\resizebox{\textwidth}{!}{
\begin{tabular}{l l l c c c c}
\toprule
\textbf{Predictor} & \textbf{Algorithm} & \textbf{Feature/Embedding Type} & \textbf{Accuracy} & \textbf{Sensitivity} & \textbf{Specificity} & \textbf{MCC} \\
\midrule
iBitter-SCM~\cite{charoenkwan2020ibitter} & Scoring Card Method (SCM) & Propensity scores of amino acids and dipeptides & 84.0 & 84.0 & 84.0 & 0.69 \\
BERT4Bitter~\cite{charoenkwan2021bert4bitter} & BERT + Bi-LSTM & BERT embeddings & 92.0 & 94.0 & 91.0 & 0.84 \\
iBitter-Fuse~\cite{charoenkwan2021ibitter} & SVM & Composition + Physicochemical properties & 93.0 & 94.0 & 92.0 & 0.86 \\
iBitter-DRLF~\cite{jiang2022identify} & LightGBM & SSA, UniRep, and BiLSTM embeddings & 94.0 & 92.0 & 98.0 & 0.89 \\
UniDL4BioPep~\cite{du2023unidl4biopep} & CNN (shallow, 8-layer) & ESM-2 embeddings (320-dim) & 93.8 & 92.4 & 95.2 & 0.87 \\
Bitter-RF~\cite{zhang2023bitter} & Random Forest & Physicochemical sequence features & 94.0 & 94.0 & 94.0 & 0.88 \\
iBitter-GRE~\cite{zhang2023bitter} & Stacking Ensemble & ESM-2 embeddings + Biochemical Descriptors & 96.1 & 98.4 & 93.8 & 0.92 \\
\bottomrule
\end{tabular}}
\end{table*}

Further extending this line of research, \textit{iBitter-DRLF}~\cite{jiang2022identify} incorporated deep representation learning techniques to develop an ML-based model for bitter peptide identification. This model leveraged two types of peptide sequence-based feature extraction methods, resulting in enhanced classification performance compared to traditional non-deep learning methods. Experimental evaluations showed that iBitter-DRLF achieved an independent test accuracy of 0.944, specificity of 0.977, MCC of 0.889, and an Area Under the Receiver Operating Characteristic (auROC) of 0.977, making it one of the most advanced predictors currently available. Additionally, Unified Manifold Approximation and Projection (UMAP) dimensionality reduction was utilized to explore feature representation strategies, further strengthening the model’s interpretability. Nonetheless, the reliance on limited types of deep representations and the absence of ensemble learning techniques may restrict the model’s ability to fully capture complementary feature information across diverse embedding schemes.

Extending recent advancements in computational approaches for peptide bioactivity prediction, \textit{UniDL4BioPep}~\cite{du2023unidl4biopep} proposed a universal deep learning architecture leveraging pretrained PLMs for bioactive peptide classification. Unlike conventional methods, \textit{UniDL4BioPep} employs Evolutionary Scale Modeling (ESM-2) embeddings combined with convolutional neural networks (CNNs) to address limitations associated with traditional feature extraction techniques. ESM-2 effectively captures both residue-level and sequential context from peptide sequences, overcoming drawbacks of traditional descriptors such as dimensionality issues and sequential information loss. The method was validated across twenty datasets covering eighteen bioactivities, including bitterness prediction, and demonstrated superior or comparable performance to state-of-the-art models in fifteen of these tasks. Specifically, \textit{UniDL4BioPep} achieved a 93.8\% accuracy and an MCC of 0.875 for bitter peptide identification, surpassing previous methods like \textit{BERT4Bitter} and \textit{iBitter-Fuse}. Similar to previous NLP-based computational approaches, this method also omits physicochemical properties and compositional features of peptide sequences, potentially limiting a comprehensive understanding of the biochemical factors influencing peptide bioactivity.

Building on the early success of ESM-based embeddings in peptide classification, \textit{iBitter-GRE}~\cite{lv2025ibitter} also adopted ESM-2 (similar to \textit{UniDL4BioPep}) in combination with handcrafted biochemical descriptors to construct a stacking ensemble model for bitter peptide prediction. Specifically, the study used a 6-layer, 8M parameter version of ESM-2 trained on UniProt, alongside seven manually engineered features including molecular weight, hydrophobicity, polarity, isoelectric point, amino acid composition, transition frequency, and amino acid distribution. These features were fused and subjected to dimensionality reduction using recursive feature elimination with cross-validation (RFECV). The ensemble framework comprised three fixed base classifiers (Gradient Boosting, Random Forest, and Extra Trees) and employed Logistic Regression as the meta-classifier. This configuration achieved strong performance, reporting an accuracy of 96.1\% and an MCC of 0.923. While effective, the iBitter-GRE framework exhibits several limitations. First, the reliance on a fixed set of base classifiers without exploring a broader space of learner-feature combinations limits the flexibility and potential optimization of the ensemble. Second, the early fusion of ESM-derived embeddings with physicochemical descriptors does not account for the distinct predictive contributions of each feature type, potentially introducing redundancy or feature dilution. Third, although the model incorporates a subset of biochemical descriptors, it omits several informative sequence-level representations, such as Amino Acid Entropy (AAE), Grouped Tripeptide Composition (GTC), and Composition-Transition-Distribution (CTD), that could enrich the model’s understanding of peptide structure-function relationships.

To advance the field beyond these limitations, we propose a novel stacking-based ensemble learning framework that combines diverse peptide representations and machine learning classifiers into a unified meta-learning pipeline (as illustrated in Fig.~\ref{fig:pipeline}). Our approach integrates embeddings derived from Protein Language Models (PLMs) such as ESM with handcrafted physicochemical and compositional descriptors. This multi-view feature strategy not only captures contextual and structural nuances of peptide sequences but also incorporates domain-specific biochemical characteristics associated with bitterness perception. Unlike previous models, such as iBitter-GRE, which also utilizes ESM-2 embeddings and traditional descriptors but relies on fixed base classifier configuration, our framework systematically constructs 56 base learners across multiple encoding-classifier combinations, then selects the top-performing ones to form a refined meta-dataset. A Logistic Regression model, chosen for its balance of robustness and empirical performance, is used as the final meta-learner. This design enhances robustness, avoids overfitting, and provides a more generalizable prediction architecture for bitter peptide identification.

\section{\textbf{Materials and Methods}}
\label{sec:materialsandmethods}

\subsection{\textbf{Dataset}}

A well-curated benchmark dataset forms the foundation of any reliable and generalizable machine learning model, particularly in the domain of bitter peptide prediction. To ensure robustness, reproducibility, and fair evaluation, we adopted the BTP640 dataset, a widely accepted benchmark dataset that has been frequently used in prior research~\cite{charoenkwan2020ibitter}. This dataset comprises 320 experimentally validated bitter peptides and 320 non-bitter peptides, making it balanced and suitable for binary classification tasks. The bitter peptide sequences were meticulously collected from multiple peer-reviewed studies, ensuring that only peptides with strong experimental validation were included~\cite{drewnowski2000bitter}~\cite{kim2006quantitative}~\cite{wu2007quantitative}~\cite{yin2010studying}~\cite{tong2008novel}~\cite{lin2008new}~\cite{liang2009using}~\cite{xu2019quantitative}. 

To maintain the integrity and quality of the dataset, several filtering steps were applied. Peptides containing ambiguous amino acid residues such as X, B, U, and Z were excluded, as these symbols often indicate uncertainty or uncommon modifications in peptide sequences. Furthermore, duplicate sequences were removed to avoid data redundancy and overfitting~\cite{vasylenko2015scmpsp}~\cite{vasylenko2016scmbyk}~\cite{charoenkwan2013scmcrys}~\cite{liou2015scmmtp}. This rigorous curation process ensures that the dataset reflects biologically meaningful and non-redundant examples of bitter peptides. On the other hand, experimentally validated non-bitter peptides are relatively limited in number, as research interest and resources have traditionally focused more on identifying functional or bioactive peptides. To address this scarcity, the negative dataset was constructed following established practices reported in previous studies~\cite{zheng2018bitter}. Specifically, non-bitter peptides were randomly selected from the BIOPEP database~\cite{minkiewicz2008biopep}, which is widely recognized as a comprehensive and reliable source of peptide sequences. The use of the BIOPEP database for generating non-bitter counterparts has been validated in various computational peptide prediction studies~\cite{gautam2013silico}~\cite{kumar2015silico}~\cite{tyagi2013silico}. By carefully balancing the dataset with equal numbers of bitter and non-bitter sequences, we ensured that our classification model would not be biased toward either class.

To facilitate fair and unbiased model training and evaluation, the dataset was randomly divided into training and independent test subsets using an 8:2 ratio, a widely accepted convention in ML-based peptide classification research~\cite{charoenkwan2020ibitter}~\cite{charoenkwan2020idppiv}~\cite{charoenkwan2020iumami}~\cite{charoenkwan2020ibitter}~\cite{wei2021computational}. The training set, referred to as BTP-CV, includes 256 bitter peptides and 256 non-bitter peptides, while the independent test set, termed BTP-TS, contains 64 bitter and 64 non-bitter peptides. This stratified sampling approach preserves class balance across both subsets, thereby ensuring consistent evaluation of model performance.

The BTP640 dataset has been widely adopted in the literature, including in the development of models such as iBitter-SCM and BERT4Bitter~\cite{charoenkwan2020ibitter, charoenkwan2021bert4bitter}. Its public availability promotes transparency, reproducibility, and fair benchmarking across studies. The dataset and corresponding source code can be accessed at \url{https://github.com/Shoombuatong/Dataset-Code/tree/master/iBitter} and \url{http://pmlab.pythonanywhere.com/BERT4Bitter}. By leveraging this standardized benchmark, our study ensures rigorous evaluation and enables direct comparison with existing state-of-the-art methods in bitter peptide prediction.

To further mitigate potential information leakage and ensure a fairer evaluation, we also performed an additional experiment where sequences sharing high identity across training and testing sets were explicitly removed. Specifically, we applied pairwise global alignment and filtered out any peptides with greater than or equal to 80\% sequence identity, both within and across the train-test boundary. The resulting filtered dataset included 428 training and 86 testing sequences, with a slight class imbalance. A detailed description of this process and its effect on model performance is provided in ~\ref{appendix:threshold}. These results confirm that our model maintains robust predictive capability even under stricter similarity constraints.

\subsection{\textbf{Feature Representation}}

Given a peptide sequence \( P \), it can be represented as:

\[
P = p_1 p_2 p_3 \dots p_N
\]

where \( p_i \) denotes the \( i \)-th residue in the sequence \( P \), and \( N \) is the total length of the peptide. Each residue \( p_i \) is selected from the standard set of 20 natural amino acids: A, C, D, E, F, G, H, I, K, L, M, N, P, Q, R, S, T, V, W, and Y. To investigate the influence of different sequence attributes on bitter peptide classification, we employed a range of feature encoding schemes. These include NLP-based embeddings from ESM, composition-based descriptors such as DPC, position-specific encodings like Amino Acid Entropy (AAE), and several physicochemical property-based descriptors including Amino Acid Index (AAI), Global Tri-Peptide Composition (GTPC), and Composition-Transition-Distribution (CTD).

The primary objective of using such a diverse array of features was to construct a comprehensive representation of peptide sequences. ESM embeddings offer deep representation learning capabilities by capturing rich evolutionary and contextual information from protein sequences. These embeddings have shown superior performance in modeling sequence dependencies and capturing high-level biological signals. In contrast, DPC and other composition-based encodings quantify the occurrence frequency of individual or paired amino acids, providing a straightforward statistical summary of the sequence.

In addition to global features, we incorporated position-specific encodings to capture the structural and functional significance of terminal residues. Specifically, we extracted features from the first five residues at the N-terminus (NT5) and the last five residues at the C-terminus (CT5). These terminal regions often play a pivotal role in determining peptide bioactivity, as they contribute to molecular stability, receptor binding, and overall structural conformation~\cite{chung2020characterization}~\cite{chaudhary2016web}. Encoding these terminal regions thus provides valuable local information that complements global sequence features and enhances predictive accuracy. To further enrich the feature space, we included physicochemical property-based descriptors. These features consider intrinsic chemical properties such as hydrophobicity, charge, polarity, and molecular volume, which are directly associated with bitterness perception and biological function. The integration of physicochemical descriptors introduces a layer of biochemical context that purely sequence-based features may lack.

\begin{figure*}[!t]
    \centering
    \includegraphics[width=0.8\textwidth]{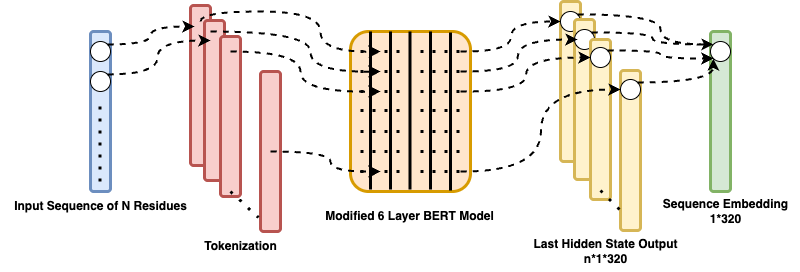}
    \caption{Architecture of the ESM-2 model used for generating peptide embeddings. The input amino acid sequence is tokenized and passed through multiple transformer layers. Final embeddings are extracted from the last layer (layer 6 in esm2\_t6\_8M\_UR50D), resulting in a 320-dimensional representation that captures evolutionary and contextual information relevant for bitter peptide prediction.}
    \label{fig:esmarchitecture}
\end{figure*}

All feature encoding techniques were systematically integrated to assess their individual and complementary roles in bitter peptide recognition. This comprehensive representation strategy enables the model to capture both high-level contextual abstractions from language models and low-level biochemical and positional information inherent in peptide sequences. By combining deep embeddings with handcrafted descriptors, the approach ensures that diverse facets of peptide characteristics are effectively captured. The results of these encoding strategies, along with their comparative performance across classifiers, are presented in the following sections. Through this multi-perspective evaluation, we aim to develop a robust, generalizable, and interpretable model for accurate bitter peptide prediction.

\subsubsection{\textbf{Evolutionary Scale Modeling (ESM) Embeddings}}

ESM is a Language Model (LM) project initiated by FAIR in $2019$ with the goal of enhancing protein and peptide sequence representations using evolutionary information \cite{lin2023evolutionary}. The most recent update, ESM-2, was trained on the UR50/D2021\_04 dataset and has demonstrated superior performance in a variety of structural prediction tasks. ESM-$2$ contains multiple model variants, ranging from $48$ layers with $15$ billion parameters to smaller versions with $6$ layers and $8$ million parameters. Given the size of the bitter peptide datasets in this study, the esm2\_t6\_8M\_UR50D variant with $320$ output embedding dimensions was selected to generate peptide embeddings. This choice helps simplify the model architecture and mitigate the curse of dimensionality during model development.

For our analysis, each peptide sequence was input into the pretrained ESM-2 model to produce a $1×320$-dimensional vector. These embeddings were extracted from the last layer of the model, specifically layer 6, which provides the most relevant sequence information for bioactivity recognition, including bitter peptide identification. The training and test datasets were split following standard procedures, using the same datasets as those used in state-of-the-art models for consistency. The embeddings generated by ESM-2 were then normalized using min-max normalization based on the training dataset, scaling the features within the range of [0, 1]. The test dataset underwent normalization using the minimum and maximum values derived from the training set to ensure consistent feature scaling during the model evaluation process. The architecture of ESM embedding model is illustrated in Fig.~\ref{fig:esmarchitecture}.

In order to assess the effectiveness of ESM-2 in capturing relevant features for peptide bioactivity prediction, we utilized Uniform Manifold Approximation and Projection (UMAP) to visualize the high-dimensional embeddings in a two-dimensional space \cite{mcinnes2018umap}. Additionally, t-Distributed Stochastic Neighbor Embedding (t-SNE) was employed to further explore the structure and separability of the embeddings based on their bioactivity class \cite{hinton2008visualizing}. These visualization techniques demonstrated that ESM embeddings provide a rich and informative representation of the peptide sequences, crucial for identifying bitter peptides.

By integrating ESM-2 embeddings, our feature representation model can leverage evolutionary sequence information alongside traditional sequence descriptors, leading to more accurate predictions of bitter peptides. These embeddings play a central role in the performance of our machine learning-based predictor, as detailed in the subsequent sections.

\subsubsection{\textbf{Dipeptide Composition (DPC)}}

DPC is a widely used feature encoding technique that captures the local relationship between adjacent amino acid residues in a peptide sequence. It is represented as a 400-dimensional vector, where each dimension corresponds to the normalized frequency of a specific dipeptide combination out of the 20\(\times\)20 possible amino acid pairs. Given a peptide sequence, the DPC encoding describes the frequencies of all possible 400 dipeptide combinations within the sequence. The calculation method is defined as follows:

\begin{equation}
D(r, s) = \frac{N_{rs}}{N - 1}, \quad r, s \in \{A, C, D, \dots, Y\}
\end{equation}

where \( N_{rs} \) denotes the number of occurrences of the dipeptide formed by amino acid types \( r \) and \( s \), and \( N \) is the total length of the peptide. The denominator \( N - 1 \) reflects the total number of adjacent amino acid pairs (dipeptides) in a sequence of length \( N \).

The DPC feature extraction process involves scanning the peptide sequence to count occurrences of each of the 400 dipeptides and then normalizing these counts to obtain relative frequencies. This normalization ensures that the resulting feature vector captures the proportional representation of each dipeptide, making it robust to variation in sequence length. DPC is particularly effective for capturing local sequential patterns, which are essential for recognizing functional properties such as bitterness. These features are integrated into the machine learning framework to support accurate classification of bitter peptides.

\subsubsection{\textbf{AAE}}

AAE is a position-based feature that quantifies the nonrandom distribution of each amino acid in a peptide sequence. AAE captures the variability and disorder of amino acid occurrences along the peptide chain, which is crucial for understanding peptide structure and functionality. The entropy is calculated based on the concept of molecular information entropy, where a lower entropy value indicates a more ordered sequence, while a higher value suggests more randomness in the distribution of amino acids.

For a peptide sequence \( P \) of length \( p \), the entropy value for each amino acid \( A \) is given by the following equation:

\begin{equation}
AAE_A = \sum_{i=1}^{n} \left( \frac{s_i - s_{i-1}}{p} \right) \log_2 \left( \frac{s_i - s_{i-1}}{p} \right)
\end{equation}

where \( p \) represents the length of the peptide sequence \( P \), \( n \) denotes the number of occurrences of the amino acid \( A \) in the peptide, and \( s_1, s_2, \dots, s_n \) are the positions of amino acid \( A \) within the peptide. The position indices are defined such that \( s_0 = 0 \) and \( s_{n+1} = n + 1 \), marking the boundaries of the peptide sequence. 

In practice, AAE is calculated not only for the full peptide sequence but also for its N-terminal (NT5) and C-terminal (CT5) subsequences, which are often critical for understanding peptide folding, bioactivity, and receptor interactions. The resulting AAE values for the full sequence, NT5, and CT5 are then combined into a single feature vector, typically consisting of 60 dimensions (20 amino acids × 3 regions: full, NT5, CT5). By incorporating AAE into the feature set, we capture critical structural information about the peptide's amino acid distribution, which plays a significant role in identifying functional characteristics such as bitterness in peptides. This entropy-based feature is integrated into machine learning models to enhance their ability to predict peptide bioactivity, particularly for applications involving bitter peptide recognition.

\subsubsection{\textbf{Binary Profile-based Encoding for N- and C-terminal residues (BPNC)}}

In Binary Profile-based Numerical Coding (BPNC), each amino acid in a peptide sequence is represented using a 20-dimensional binary vector, where each position corresponds to one of the 20 standard amino acids. The presence of a specific amino acid is denoted by a value of 1 at the respective position, while all other positions are set to 0. This encoding scheme captures positional specificity and simplifies sequence modeling for machine learning tasks. Table~\ref{tab:bpnc-example} illustrates the binary vector representations for three amino acids, Alanine (A), Cysteine (C), and Tyrosine (Y).

\begin{table}[h!]
\centering
\caption{Binary Profile of Amino Acids in BPNC Representation}
\label{tab:bpnc-example}
\resizebox{\columnwidth}{!}{
\begin{tabular}{c|l}
\toprule
\textbf{Amino Acid} & \textbf{20-Dimensional Binary Vector} \\
\midrule
A & (1, 0, 0, 0, 0, 0, 0, 0, 0, 0, 0, 0, 0, 0, 0, 0, 0, 0, 0, 0) \\
C & (0, 1, 0, 0, 0, 0, 0, 0, 0, 0, 0, 0, 0, 0, 0, 0, 0, 0, 0, 0) \\
Y & (0, 0, 0, 0, 0, 0, 0, 0, 0, 0, 0, 0, 0, 0, 0, 0, 0, 0, 0, 1) \\
\bottomrule
\end{tabular}}
\end{table}

In this study, BPNC encoding is applied specifically to the first five residues (N-terminal, NT5) and the last five residues (C-terminal, CT5) of each peptide sequence. This results in a 200-dimensional vector for each peptide, as the NT5 and CT5 sequences are encoded separately, each contributing 100 dimensions. The encoded vectors provide a representation of the peptide’s terminal residues, capturing their unique amino acid compositions in a binary format. The binary profile-based approach has been widely utilized in peptide analysis due to its simplicity and effectiveness in representing the presence or absence of specific amino acids. By focusing on the NT5 and CT5 sequences, BPNC encoding emphasizes the critical role of terminal residues in peptide function, bioactivity, and interactions with receptors.

The encoding process takes each peptide sequence and split it into corresponding NT5 and CT5 subsequences, and each amino acid in these regions is mapped to a binary vector. The final encoded feature for each peptide is a concatenation of these two subsequences, resulting in a 200-dimensional vector. These BPNC vectors are then integrated into the feature set used for training machine learning models to predict bioactivity, particularly for identifying bitter peptides. BPNC encoding offers a straightforward yet powerful means of capturing essential sequence information, particularly from the N- and C-terminal regions of peptides, which are often critical for their biological activity and recognition by receptors.

\begin{figure*}[h]
    \centering
    \includegraphics[width=\linewidth]{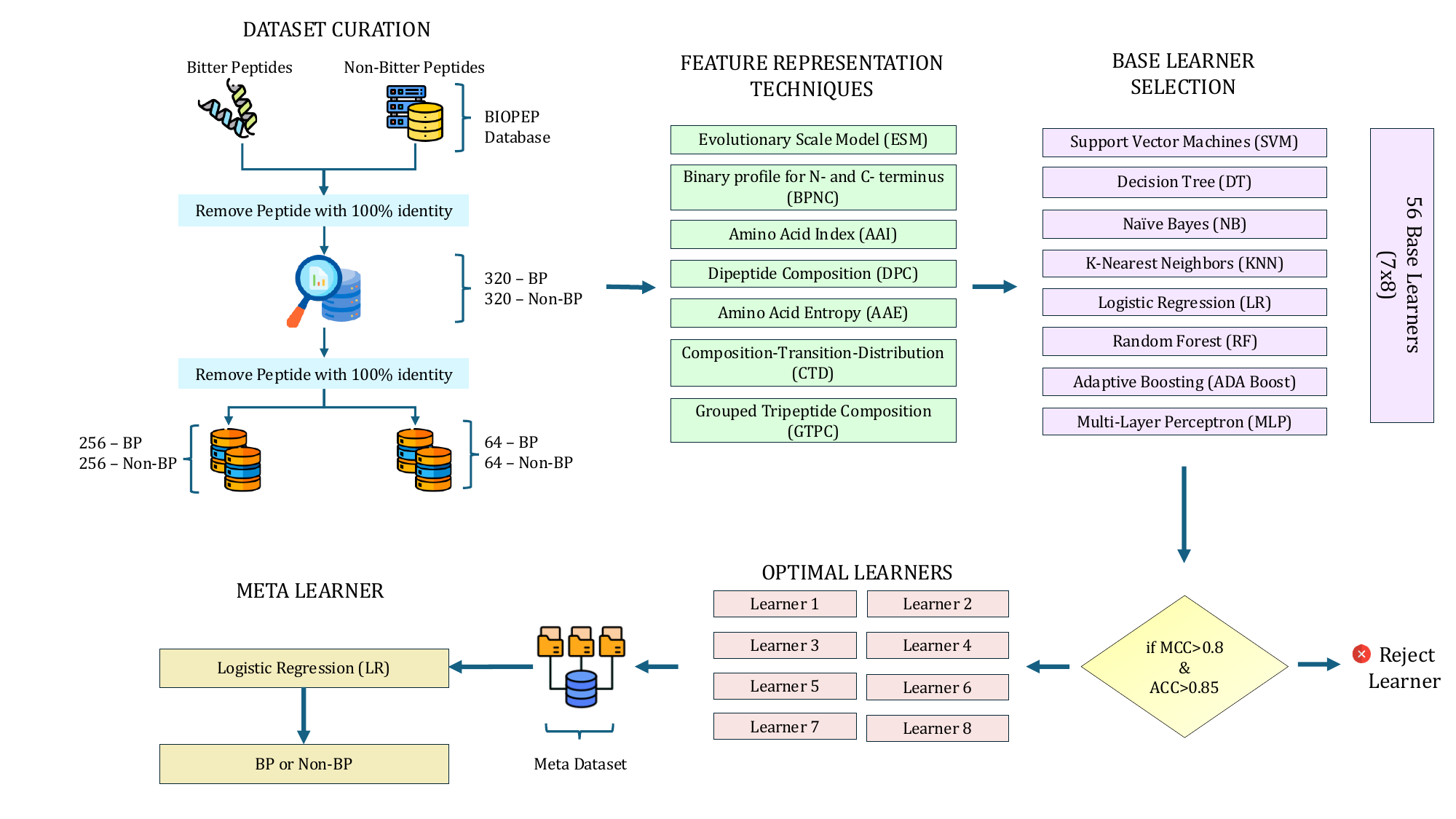}
    \caption{Overview of the proposed bitter peptide prediction pipeline. The workflow begins with \textbf{Dataset Curation}, where experimentally validated bitter peptides and randomly selected non-bitter peptides from the BIOPEP database are filtered for 100\% identity and split into training and test sets. In the \textbf{Feature Representation} stage, seven distinct encoding techniques are applied to capture both contextual and physicochemical properties of peptide sequences: ESM, BPNC, AAI, DPC, AAE, CTD, and GTPC. In the \textbf{Base Learner Selection} phase, each of the seven feature types is used to train eight machine learning classifiers (SVM, DT, NB, KNN, LR, RF, AdaBoost, and MLP), resulting in 56 base models. Models not satisfying performance thresholds (MCC $>$ 0.80 and ACC $>$ 0.9) are discarded. The \textbf{Optimal Learners} that pass the threshold contribute their predicted probabilities to form a meta-dataset, which is then processed by a \textbf{Meta Learner} (Logistic Regression) for final binary classification (Bitter Peptide or Non-Bitter Peptide). This stacked ensemble approach integrates multiple feature spaces and classifier outputs to enhance prediction robustness and accuracy.}
    \label{fig:pipeline}
\end{figure*}

\subsubsection{\textbf{Physicochemical Property-Based Features}}

Physicochemical property-based features are crucial for capturing the chemical characteristics and structural properties of peptides, which play an important role in their bioactivity. In this study, we employ three key physicochemical feature encodings: AAI, GTPC, and CTD. These features provide detailed information about amino acid distributions, chemical properties, and sequence motifs, which are essential for accurately predicting peptide bioactivity, including bitterness.

\paragraph{\textbf{AAI}}

The AAI consists of $12$ properties from the AAindex database, which represent various physicochemical characteristics, such as hydrophobicity, steric parameters, and solvation. For properties like hydrophobicity, hydrophilicity, steric parameter, solvation, and others, the average AAindex values of all amino acids in the full, NT5, and CT5 sequences are used. For properties such as hydrogen bonding, net charge, and molecular weight, the sum of the AAindex values for all amino acids in the sequence is used. AAI is encoded as a 36-dimensional vector, where each dimension corresponds to a specific AAindex property. This feature encoding technique provides important chemical information about the peptide's amino acid composition, contributing to the model's ability to predict bioactive properties.

\paragraph{\textbf{GTPC}}

GTPC categorizes amino acids into five groups based on their physicochemical properties: aliphatic, aromatic, positive charge, negative charge, and uncharged. GTPC is calculated by determining the frequency of tri-peptides formed by combinations of these amino acid groups in the full, NT5, and CT5 sequences. The resulting vector has 125 dimensions, each corresponding to a specific combination of amino acid groups in a tri-peptide. This encoding captures the interactions between amino acids in the sequence and is particularly useful for identifying sequence motifs that may influence bioactivity, such as bitterness.

\paragraph{\textbf{CTD}}

The CTD feature captures the distribution patterns of amino acids according to specific physicochemical properties. It consists of a 147-dimensional vector, including $21$ dimensions for the composition (C), $21$ dimensions for the transition (T), and $105$ dimensions for the distribution (D) of amino acids within the peptide sequence. These dimensions describe how amino acids with different chemical properties are distributed along the peptide chain. The CTD feature provides a comprehensive view of the peptide’s chemical environment, enabling better understanding of its structural and functional properties, which is particularly relevant for identifying bioactive peptides.

Together, these physicochemical property-based features, AAI, GTPC, and CTD, form a robust feature representation for peptide sequences. By integrating these features, we capture important chemical and structural properties that contribute to accurate peptide bioactivity predictions, enhancing the overall performance of our machine learning models.

\subsection{\textbf{Base Learners and Meta Learners}}

In this study, we constructed a powerful predictive model by combining multiple embeddings and classifiers, resulting in a total of $56$ base learners. These base learners were generated by using seven different embeddings and eight distinct classifiers, as shown in Fig.~\ref{fig:pipeline}. The embeddings used in this work include ESM, BPNC, DPC, AAE, AAI, GTPC, and CTD, while the classifiers consist of SVM, Decision Tree (DT), Naive Bayes (NB), K-Nearest Neighbors (KNN), Logistic Regression (LR), Random Forest (RF), Adaptive Boosting (AdaBoost), and Multilayer Perceptron (MLP). Each combination of an embedding and classifier was treated as an individual base learner, leading to a total of 56 models. To optimize the parameters of these models, the $56$ base learners were trained using a $10$-fold cross-validation setting. This process allowed us to find the optimal hyperparameters for each model, ensuring that the best possible configuration was used for training. 

After training all $56$ base learners on the training set, we applied a selection criterion to identify the top-performing models. The selection was based on two key performance metrics: MCC greater than $0.8$ and accuracy higher than $90$\%. This rigorous selection process ensured that only the most reliable and accurate models were included in the final ensemble. From this filtering process, the top eight models, comprising diverse classifiers such as SVM, RF, KNN, and others, were chosen to participate in the meta-learning phase. For each peptide sample in the training set, these eight models output a class probability indicating the likelihood of the sample being bitter or non-bitter. Specifically, each base learner returned a probability score between $0$ and $1$, rather than a hard label. These outputs were concatenated to form an $8$-dimensional probability vector for every peptide. This set of probability vectors across all samples constituted the meta dataset, which was then used as input to the meta learner.

The meta learner, positioned in the second layer of the stacking framework, received the 8-dimensional probability vectors generated by the selected top-performing base learners. Each element in this vector corresponded to the predicted probability (soft output) from one of the eight classifiers, such as SVM, RF, KNN, and others, indicating the likelihood of a peptide being bitter or non-bitter. Rather than making individual hard decisions, these learners provided nuanced confidence scores that collectively formed the meta dataset. The meta learner, implemented as a Logistic Regression (LR) model, was trained on this meta dataset to learn the optimal way to combine the predictions of the base learners. By assigning appropriate weights to each input probability, the LR model effectively captured the complementary strengths of the underlying classifiers. This ensemble strategy enhanced both the predictive accuracy and the robustness of the final model.

This two-tier architecture, comprising a diverse ensemble of base learners followed by a meta learner, enabled the system to capture complex and heterogeneous patterns within peptide sequences. The final classification was derived from the collective judgment of the most reliable models, making the framework highly effective for distinguishing between bitter and non-bitter peptides (illustrated in Fig.~\ref{fig:pipeline}).

\begin{figure*}[h!]
    \centering
    \includegraphics[width=\textwidth]{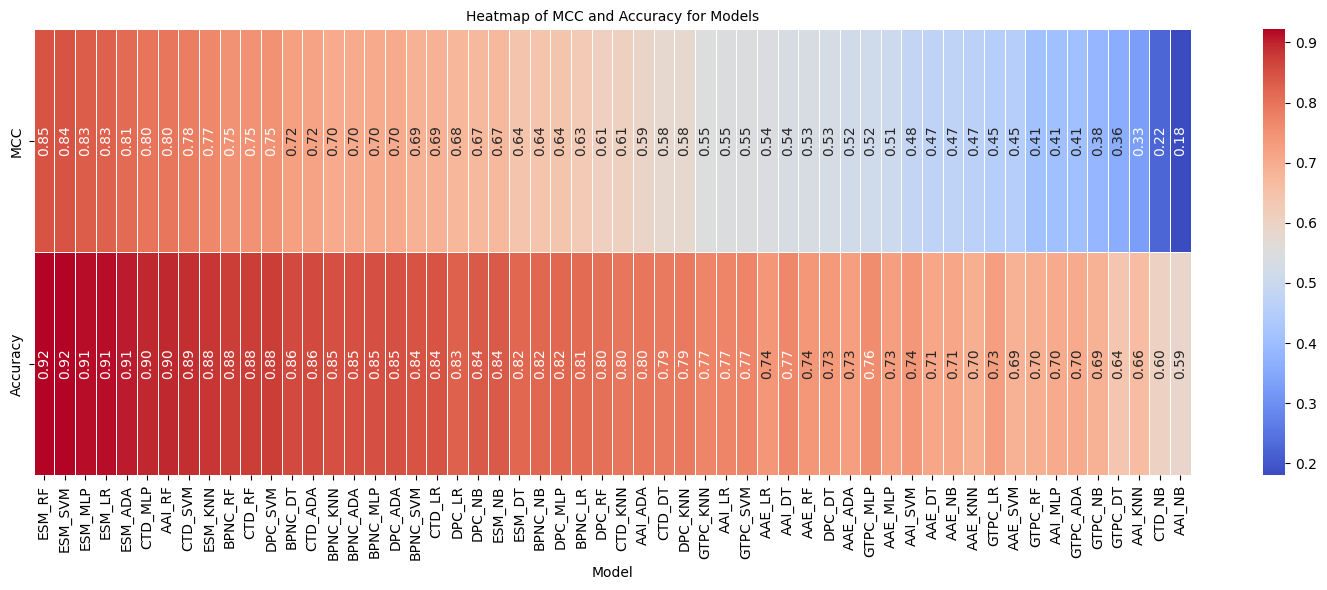}
    \caption{Heatmap of MCC and Accuracy scores across 56 base learners combining different feature extraction techniques and classifiers.}
    \label{fig:heatmap}
\end{figure*}

\subsection{\textbf{Evaluation Measures}}

The performance of the model was evaluated using several standard metrics, including Accuracy (ACC), Sensitivity (Sn), Specificity (Sp), MCC, and AUROC. These metrics were calculated as follows:

\begin{equation}
\scriptsize
\text{ACC} = \frac{\text{TP} + \text{TN}}{\text{TP} + \text{TN} + \text{FP} + \text{FN}} \quad
\end{equation}

\begin{equation}
\scriptsize
\text{Sn} = \frac{\text{TP}}{\text{TP} + \text{FN}} \quad
\end{equation}

\begin{equation}
\scriptsize
\text{Sp} = \frac{\text{TN}}{\text{TN} + \text{FP}} \quad 
\end{equation}

\begin{equation}
\scriptsize
\text{MCC} = \frac{\text{TP} \times \text{TN} - \text{FP} \times \text{FN}}{\sqrt{(\text{TP} + \text{FP})(\text{TP} + \text{FN})(\text{TN} + \text{FP})(\text{TN} + \text{FN})}} \quad
\end{equation}

\noindent
These metrics are widely used in peptide classification tasks~\cite{charoenkwan2021ibitter, jiang2022identify, zhang2023bitter}. Here, TP, TN, FP, and FN denote true positives, true negatives, false positives, and false negatives, respectively. ACC measures overall correctness, while Sn and Sp assess the model’s ability to correctly identify bitter and non-bitter peptides. Although MCC is particularly advantageous for imbalanced datasets, it remains informative in balanced settings due to its consideration of all confusion matrix components, providing a more nuanced evaluation than accuracy alone. AUROC, being threshold-independent, captures the model’s discriminative ability across varying classification thresholds. Together, these metrics provide a comprehensive and reliable assessment of the model’s performance.

\section{\textbf{Results and Discussions}}
\label{sec:resultsanddiscussions}

\subsection{\textbf{Performance Evaluation of Base Learners}}

To evaluate the effectiveness of individual models trained on diverse feature representations, a comprehensive comparison of $56$ base learners was performed using two key performance metrics: MCC and Accuracy. Fig.~\ref{fig:heatmap} presents a heatmap summarizing the performance of each base learner across different feature types and classifiers. As observed in Fig.~\ref{fig:heatmap}, models based on the ESM combined with ensemble classifiers such as RF, SVM, and MLP consistently outperformed other combinations. The highest-performing model, \texttt{ESM\_RF}, achieved an MCC of $0.85$ and an accuracy of $92$\%. In contrast, models built using the AAI and GTPC features generally demonstrated lower performance across both metrics.

\begin{figure}[h!]
    \centering
    \includegraphics[width=0.4\textwidth]{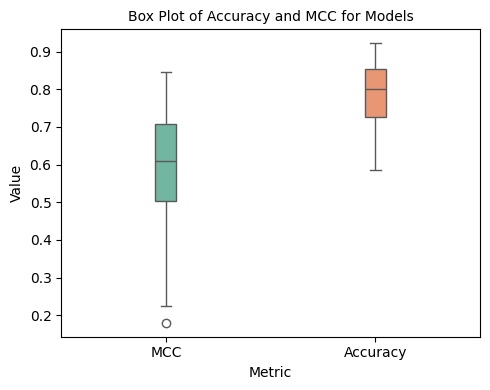}
    \caption{Box plot showing the distribution of MCC and Accuracy values across all base learners.}
    \label{fig:boxplot}
\end{figure}

To further illustrate the distribution and variability in performance, a box plot of MCC and Accuracy values across all models is shown in Fig.~\ref{fig:boxplot}. The median accuracy across models remains relatively high, with a narrow inter-quartile range, whereas MCC exhibits a broader spread with a few models under-performing significantly. This variability in MCC, despite the balanced dataset, can be attributed to its sensitivity to both false positives and false negatives, which are not equally captured by accuracy. Even with balanced class distribution, models that misclassify difficult or ambiguous cases may achieve high accuracy but lower MCC, revealing discrepancies in their overall predictive reliability.

This comparative analysis guided the selection of top-performing base learners for inclusion in the meta-dataset, retaining only those models that achieved an MCC greater than $0.8$ and an accuracy above $90$\% for the stacking ensemble.

\subsection{\textbf{Identification of Optimal Base Learners for Meta-Modeling}}

Building on the results of the comparative analysis, the most effective base learners were selected to construct the final stacked ensemble. A total of eight models met the performance criteria and were retained for meta-modeling. These top-performing models span multiple feature representation techniques, including:

\begin{itemize}
    \item \textbf{ESM}: \texttt{ESM\_RF}, \texttt{ESM\_SVM}, \texttt{ESM\_MLP}, \texttt{ESM\_LR}, \texttt{ESM\_ADA}
    \item \textbf{CTD}: \texttt{CTD\_MLP}, \texttt{CTD\_SVM}
    \item \textbf{AAI}: \texttt{AAI\_RF}
\end{itemize}

\noindent Notably, while ESM-derived models dominated the top tier, the inclusion of CTD and AAI-based models indicates that complementary information from alternative descriptors contributes meaningfully to overall predictive performance. To further validate this, we also constructed a restricted meta-learner using only the five ESM-based models termed ESM\_Stack. Although this ESM-only ensemble achieved competitive results (shown in Table~\ref{tab:test-results}), its performance remained consistently lower than our proposed model, which integrates CTD and AAI-based learners and achieves an MCC of $0.922$ and accuracy of $96.1$\%. These findings underscore the value of feature diversity in capturing distinct biochemical patterns relevant to bitterness prediction. Additionally, all selected models are powered by classifiers known for handling nonlinear patterns, particularly learners like RF, AdaBoost, and MLP.

To further investigate how the ensemble benefits from model diversity, we conducted a qualitative analysis of agreement patterns among the selected base learners. Notably, the five ESM-based models frequently showed high concordance in their predictions, particularly for confidently classified peptides. However, the inclusion of CTD and AAI-based learners contributed valuable complementary signals, especially in ambiguous or borderline cases where the ESM ensemble showed partial disagreement. This reinforces the merit of integrating orthogonal features and model types rather than relying solely on PLM-derived embeddings.

\begin{figure*}[!ht]
    \centering
    \includegraphics[width=\textwidth]{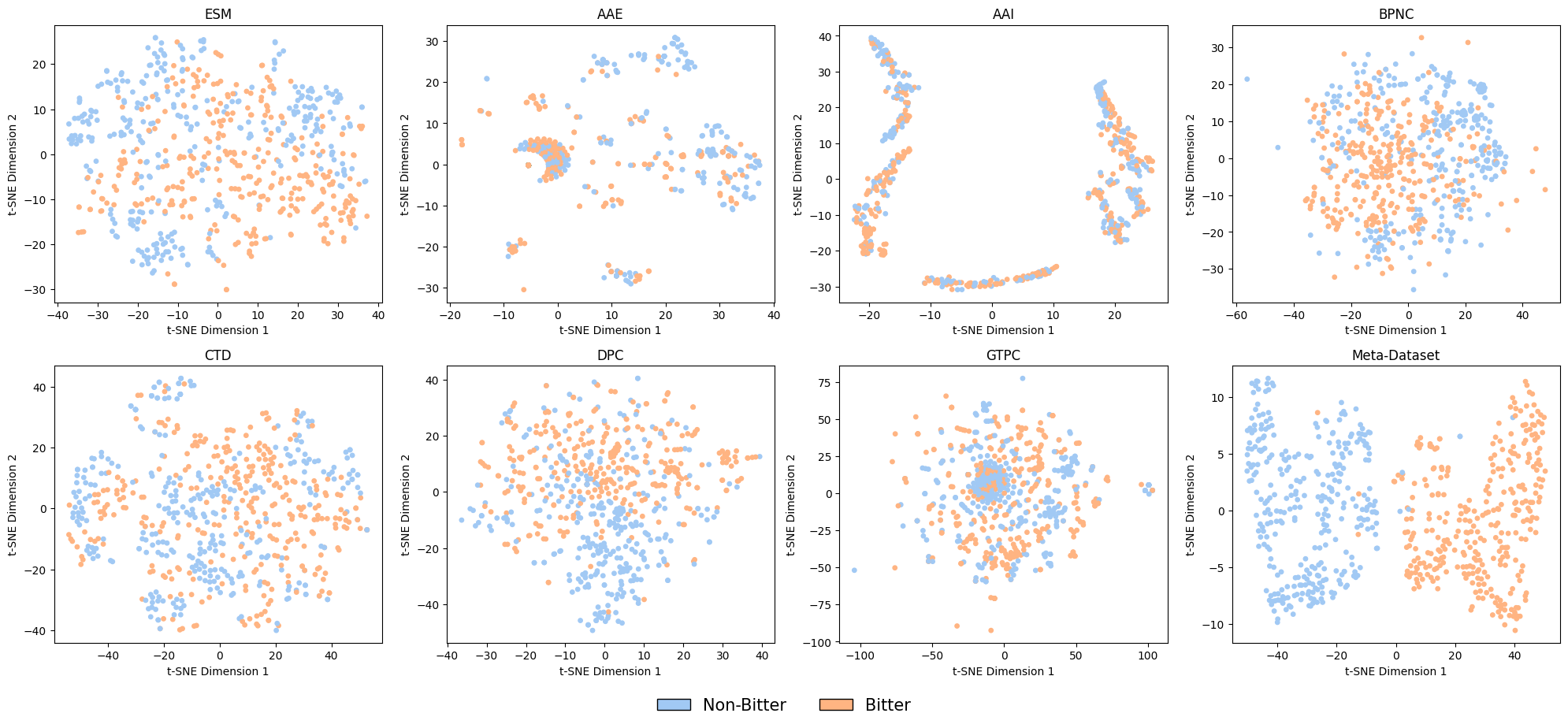}
    \caption{t-SNE visualization of bitter (orange) and non-bitter (blue) peptides across various feature representations. The final plot (bottom-right) shows separation based on the 8-dimensional meta-dataset.}
    \label{fig:tsne}
\end{figure*}

Each selected model outputs a probability score indicating the likelihood of a given peptide being bitter. These scores are then concatenated to form an $8$-dimensional vector for each sample, collectively forming the \textit{meta-dataset}. This meta-representation captures the ensemble-level consensus across high-confidence classifiers.

The meta-dataset is used to train a Logistic Regression (LR) model, serving as the final meta-learner. Among several candidate classifiers evaluated at this stage, including SVM and RF, LR consistently demonstrated superior performance in terms of both accuracy and MCC. Its ability to model linear decision boundaries while maintaining computational efficiency made it the most suitable choice for combining the soft predictions of base learners. This selection strategy not only streamlines the prediction pipeline but also reduces computational overhead by relying solely on a curated set of models during inference.

\subsection{\textbf{Performance Evaluation of the Stacked Meta-Learner}}

To improve the robustness and generalization ability of bitter peptide classification, we adopted a stacked ensemble strategy. The complete workflow is illustrated in Fig.~\ref{fig:pipeline}, detailing the stages from dataset curation and feature extraction to base learner evaluation and final meta-classification. From the pool of $56$ base learners constructed using combinations of seven feature encodings and eight classifiers, a subset of high-performing models was shortlisted to contribute to the final ensemble. Each selected model provided a class probability score, and these were concatenated into an $8$-dimensional representation per instance, collectively forming the meta-dataset. A LR model was then employed as the meta-learner due to its strong empirical performance in ensemble stacking scenarios~\cite{dvzeroski2004combining}.

Hyperparameters for the LR model were optimized via grid search, with the best configuration obtained at \texttt{penalty} = \texttt{l2} and \texttt{max\_iter} = 1500. This setting achieved a balance between regularization and convergence efficiency, particularly beneficial given the highly discriminative nature of the meta-features.

\subsubsection{\textbf{Performance Comparison with Base Learners}}

Table~\ref{tab:train-results} presents a comparative summary of the proposed meta-learner against top-performing base models using $10$-fold cross-validation. While several base learners such as \texttt{ESM\_SVM} and \texttt{ESM\_MLP} demonstrated promising performance, achieving MCC values close to $0.71$ and balanced accuracy, the stacked ensemble significantly outperformed them. The meta-learner achieved nearly perfect classification metrics, with an MCC of $0.996$, accuracy of $99.8$\%, and AUROC of $0.998$. These results underscore the effectiveness of the stacked approach in integrating diverse decision boundaries and generalizing patterns learned by individual models.

\begin{table}[!ht]
\caption{10-Fold Cross-Validation Results: Meta-Learner vs. Base Learners}
\label{tab:train-results}
\centering
\resizebox{\columnwidth}{!}{
\begin{tabular}{lccccc}
\toprule
\textbf{Model} & \textbf{Acc (\%)} & \textbf{Sn (\%)} & \textbf{Sp (\%)} & \textbf{MCC} & \textbf{AUROC} \\
\midrule
ESM\_SVM & 85.5 & 85.9 & 85.1 & 0.71 & 0.85 \\
ESM\_RF & 83.4 & 82.8 & 84.0 & 0.67 & 0.83 \\
ESM\_MLP & 83.6 & 85.1 & 82.1 & 0.67 & 0.83 \\
ESM\_LR & 83.6 & 83.9 & 83.2 & 0.67 & 0.83 \\
CTD\_MLP & 81.1 & 80.4 & 81.7 & 0.62 & 0.81 \\
ESM\_ADA & 83.0 & 80.4 & 85.6 & 0.66 & 0.83 \\
CTD\_SVM & 83.2 & 83.2 & 83.3 & 0.66 & 0.83 \\
AAI\_RF & 78.5 & 79.7 & 77.3 & 0.57 & 0.78 \\
\textbf{iBitter-Stack} & \textbf{99.8} & \textbf{100.0} & \textbf{99.6} & \textbf{0.99} & \textbf{0.99} \\
\bottomrule
\end{tabular}}
\end{table}

To assess the robustness and generalization capability of the ensemble model, we further evaluated its performance on an independent test set. As shown in Table~\ref{tab:test-results}, the meta-learner maintained a high level of predictive reliability, achieving an accuracy of $96.1$\%, MCC of $0.922$, and AUROC of $0.961$. Notably, several base learners, particularly \texttt{ESM\_RF} and \texttt{ESM\_SVM}, also demonstrated improved performance on the independent test set compared to their 10-fold cross-validation results (Table~\ref{tab:train-results}), with both achieving MCC values above $0.84$ and AUROC values above $0.92$. This performance trend suggests that while the models perform well on the independent test set, their relatively lower scores in the 10-fold cross-validation indicate potential limitations in capturing broader generalization across diverse data splits.

To better understand the discriminative power of different feature representations, including the final meta-representation, we conducted t-distributed Stochastic Neighbor Embedding (t-SNE) analysis and visualized the feature space. Fig.~\ref{fig:tsne} displays two-dimensional t-SNE projections of bitter and non-bitter peptides across various feature types. It is evident that individual features like AAE, DPC, and GTPC result in high overlap between classes, indicating limited separability. In contrast, the final 8-dimensional meta-dataset—constructed from the probability outputs of selected base learners—achieves the most distinct clustering between bitter and non-bitter samples. The clear margins and tight groupings seen in this plot suggest that the stacked representation captures a more abstract and highly discriminative decision space, further explaining the superior performance of the meta-learner across evaluation settings.

\begin{table}[!ht]
\caption{Independent Test Set Results: Meta-Learner vs. Base Learners}
\label{tab:test-results}
\centering
\resizebox{\columnwidth}{!}{
\begin{tabular}{lccccc}
\toprule
\textbf{Model} & \textbf{Acc (\%)} & \textbf{Sn (\%)} & \textbf{Sp (\%)} & \textbf{MCC} & \textbf{AUROC} \\
\midrule
ESM\_SVM & 92.2 & 92.2 & 92.2 & 0.84 & 0.92 \\
ESM\_RF & 92.2 & 89.1 & 95.3 & 0.84 & 0.92 \\
ESM\_MLP & 91.4 & 85.9 & 96.9 & 0.83 & 0.91 \\
ESM\_LR & 91.4 & 90.6 & 92.2 & 0.82 & 0.91 \\
CTD\_MLP & 89.8 & 87.5 & 92.2 & 0.79 & 0.89 \\
ESM\_ADA & 89.1 & 90.6 & 87.5 & 0.78 & 0.89 \\
CTD\_SVM & 89.1 & 85.9 & 92.2 & 0.78 & 0.89 \\
AAI\_RF & 89.8 & 90.6 & 89.1 & 0.79 & 0.89 \\
ESM\_Stack & 92.9 & 91.0 & 95.1 & 0.86 & 0.98 \\
\textbf{iBitter-Stack} & \textbf{96.1} & \textbf{95.4} & \textbf{97.2} & \textbf{0.92} & \textbf{0.98} \\
\bottomrule
\end{tabular}}
\end{table}

\subsubsection{\textbf{Comparison with State-of-the-Art Models}}

Table~\ref{tab:lit-cv-comparison} and Table~\ref{tab:lit-comparison} present a comprehensive performance comparison between the proposed model and several state-of-the-art bitter peptide classifiers, evaluated under both 10-fold cross-validation and independent test set protocols. To ensure fairness and replicability, results are compared using the same splits and evaluation standards reported in the respective studies.

\begin{table}[!ht]
\caption{Comparison with Prior State-of-the-Art Models (10-Fold Cross-Validation)}
\label{tab:lit-cv-comparison}
\centering
\resizebox{\columnwidth}{!}{
\begin{tabular}{lccccc}
\toprule
\textbf{Model} & \textbf{Acc (\%)} & \textbf{Sn (\%)} & \textbf{Sp (\%)} & \textbf{MCC} & \textbf{AUROC} \\
\midrule
iBitter-SCM~\cite{charoenkwan2020ibitter} & 87.0 & 91.0 & 83.0 & 0.75 & 0.90 \\
BERT4Bitter~\cite{charoenkwan2021bert4bitter} & 86.0 & 87.0 & 85.0 & 0.73 & 0.92 \\
iBitter-Fuse~\cite{charoenkwan2021ibitter} & 92.0 & 92.0 & 92.0 & 0.84 & 0.94 \\
iBitter-DRLF~\cite{jiang2022identify} & 89.0 & 89.0 & 89.0 & 0.78 & 0.95 \\
Bitter-RF~\cite{zhang2023bitter} & 85.0 & 86.0 & 84.0 & 0.70 & 0.93 \\
iBitter-GRE~\cite{zhang2023bitter} & 86.3 & 85.5 & 87.1 & 0.73 & 0.92 \\
\textbf{iBitter-Stack} & \textbf{99.8} & \textbf{100.0} & \textbf{99.6} & \textbf{0.99} & \textbf{0.99} \\
\bottomrule
\end{tabular}}
\end{table}

In the 10-fold cross-validation setting (Table~\ref{tab:lit-cv-comparison}), traditional models such as \textit{iBitter-SCM} and \textit{BERT4Bitter} show moderate performance, with accuracies below 88\% and MCCs under 0.75. Mid-generation models like \textit{iBitter-Fuse} and \textit{iBitter-DRLF} demonstrate more balanced predictive behavior, achieving MCCs of 0.84 and 0.78, respectively. The recent iBitter-GRE model, which integrates ESM-2 embeddings with handcrafted descriptors in a stacking framework, yields an accuracy of 86.3\%, specificity of 87.1\%, and MCC of 0.73, promising but still behind our proposed approach. In stark contrast, our model achieves near-perfect results with an accuracy of 99.8\%, MCC of 0.99, and an AUROC of 0.99. This substantial performance margin underscores the effectiveness of our stacking-based ensemble framework and its ability to generalize well across cross-validation folds.

\begin{table}[!ht]
\caption{Comparison with Prior State-of-the-Art Models (Independent Test Set)}
\label{tab:lit-comparison}
\centering
\resizebox{\columnwidth}{!}{
\begin{tabular}{lccccc}
\toprule
\textbf{Model} & \textbf{Acc (\%)} & \textbf{Sn (\%)} & \textbf{Sp (\%)} & \textbf{MCC} & \textbf{AUROC} \\
\midrule
iBitter-SCM~\cite{charoenkwan2020ibitter} & 84.0 & 84.0 & 84.0 & 0.69 & 0.90 \\
BERT4Bitter~\cite{charoenkwan2021bert4bitter} & 92.2 & 93.8 & 90.6 & 0.84 & 0.96 \\
iBitter-Fuse~\cite{charoenkwan2021ibitter} & 93.0 & 94.0 & 92.0 & 0.86 & 0.93 \\
iBitter-DRLF~\cite{jiang2022identify} & 94.0 & 92.0 & 96.9 & 0.89 & 0.97 \\
UniDL4BioPep~\cite{du2023unidl4biopep} & 93.8 & 92.4 & 95.2 & 0.87 & 0.98 \\
Bitter-RF~\cite{zhang2023bitter} & 94.0 & 94.0 & 94.0 & 0.88 & 0.98 \\
iBitter-GRE~\cite{zhang2023bitter} & 96.1 & \textbf{98.4} & 93.8 & \textbf{0.92} & 0.97 \\
\textbf{Proposed} & \textbf{96.1} & 95.4 & \textbf{97.2} & \textbf{0.92} & \textbf{0.98} \\
\bottomrule
\end{tabular}}
\end{table}

The independent test set results (Table~\ref{tab:lit-comparison}) further validate the superiority of our model in real-world settings. It achieves the highest specificity (97.2\%) among all models, along with a strong sensitivity (95.4\%), and matches the top MCC (0.92) achieved by iBitter-GRE. Although iBitter-GRE and our proposed model achieve similar MCC scores on the independent test set, iBitter-Stack introduces several key advancements in both design and generalizability. A closer examination reveals that iBitter-Stack achieves a more balanced sensitivity-specificity trade-off (95.4\% and 97.2\%, respectively), compared to iBitter-GRE (98.4\% and 93.8\%). This balance suggests better control over both false positives and false negatives, improving reliability in real-world applications. Furthermore, the advantage of iBitter-Stack becomes more evident under 10-fold cross-validation. While iBitter-GRE's performance drops notably (Accuracy = 86.3\%, MCC = 0.73), iBitter-Stack maintains high consistency (Accuracy = 99.8\%, MCC = 0.99), indicating stronger generalization across varying data splits.

Architecturally, unlike iBitter-GRE, which employs a fixed set of three base learners, iBitter-Stack systematically constructs a diverse pool of 56 base models by combining various encoding and classifier schemes. It then dynamically selects only those surpassing strict performance thresholds (MCC $>$ 0.8, Accuracy $>$ 90\%). This selective ensemble strategy enhances both robustness and adaptability. In addition, we adopt a meta-level fusion approach that leverages soft probability vectors rather than early concatenation of embeddings and handcrafted features. This reduces redundancy, encourages smoother decision boundaries, and improves interpretability. Collectively, these innovations render iBitter-Stack not only highly competitive in predictive performance but also more modular, extensible, and better suited for future deployment in bioactivity prediction pipelines.

UniDL4BioPep, another recent model leveraging ESM-2 embeddings, reports solid results (MCC: 0.87, AUROC: 0.98), but lacks feature-level diversity, unlike our approach, which explicitly combines biochemical descriptors with deep embeddings for enhanced robustness. This performance advantage stems not only from our use of pre-trained PLM like ESM, but from the structured design of our ensemble architecture (Fig.~\ref{fig:pipeline}). By constructing a broad and diverse base learner pool, selecting top performers, and synthesizing their predictions into an 8-dimensional meta-representation, the model effectively captures consensus across complementary perspectives. The final Logistic Regression meta-learner then capitalizes on these fused signals to deliver reliable and well-calibrated predictions.

The performance benefits of iBitter-Stack stem not only from leveraging pretrained language models like ESM but also from its structured ensemble architecture (Fig.~\ref{fig:pipeline}). By constructing a broad and diverse base learner pool, selecting top performers, and synthesizing their predictions into an 8-dimensional meta-representation, the model effectively captures consensus across complementary perspectives. The final Logistic Regression meta-learner then capitalizes on these fused signals to deliver reliable and well-calibrated predictions.

\begin{figure}[!ht]
    \centering
    \includegraphics[width=0.45\textwidth]{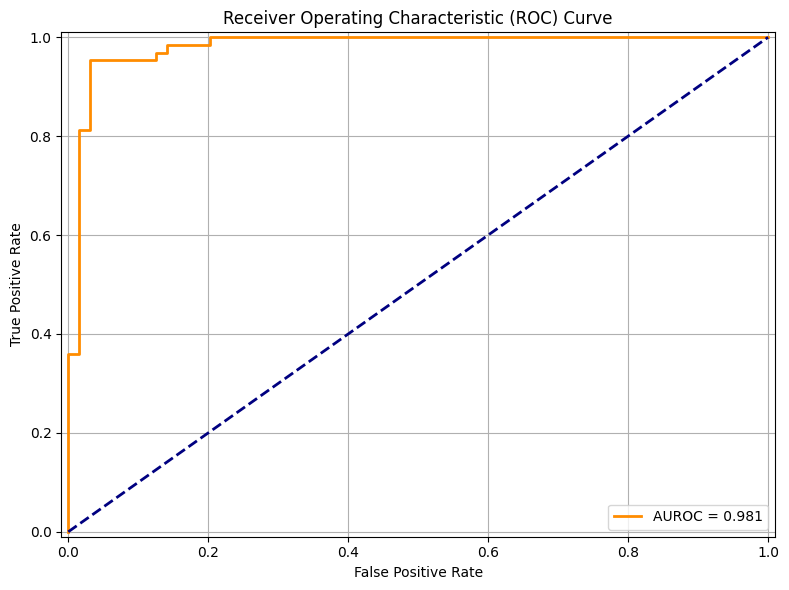}
    \caption{Receiver Operating Characteristic (ROC) Curve for the Proposed Model on the Independent Test Set. AUROC = 0.981.}
    \label{fig:roc}
\end{figure}

A further testament to our model’s reliability is illustrated by the ROC curve shown in Fig.~\ref{fig:roc}. The proposed model achieves an AUROC of 0.981 on the independent test set, demonstrating exceptional discriminatory ability across all classification thresholds. This high AUROC, on par with Bitter-RF and UniDL4BioPep, confirms that the model sustains a strong true positive rate while minimizing false positives, which is critical in peptide screening applications.

\noindent
The Matthews Correlation Coefficient (MCC), though originally highlighted for its usefulness in imbalanced settings, remains highly relevant even in balanced datasets. By considering all four confusion matrix components, true positives, true negatives, false positives, and false negatives, it offers a more reliable summary of performance than accuracy alone. The consistently high MCC of our model across both evaluation settings reflects its robustness and generalizability.

In conclusion, the proposed model advances the state of the art in bitter peptide classification by combining representational diversity, selective ensemble learning, and architectural modularity. Its strong empirical performance, paired design flexibility, positions it as a practical and powerful tool for future bioactivity prediction pipelines.

\section{\textbf{Conclusion \& Future Work}}
\label{sec:conclusion}

In this study, we proposed a stacking-based ensemble learning framework for the classification of bitter peptides. By integrating seven heterogeneous feature representations, including contextual embeddings (ESM) and handcrafted physicochemical descriptors, and combining them with eight diverse classifiers, we developed a total of $56$ base learners. A rigorous performance-based filtering strategy was employed to retain only the most effective learners, which were then used to construct an $8$-dimensional meta-dataset. This representation served as input to a logistic regression meta-learner, yielding a robust final model.

Our extensive evaluations, conducted via both $10$-fold cross-validation and independent test set analysis, revealed that the proposed ensemble consistently outperformed individual models and surpassed existing state-of-the-art predictors. Specifically, the model achieved an accuracy of $96.1$\%, an MCC of $0.922$, and an AUROC of $0.981$ on the independent test set, demonstrating strong discriminative ability and generalization.

From these results, we infer that combining multiple perspectives on peptide sequences, ranging from high-level evolutionary embeddings to fine-grained physicochemical descriptors, enables a richer, more generalizable representation of peptide behavior. The meta-learner effectively consolidates these complementary signals, leading to more reliable and well-balanced predictions. Moreover, the performance gains over strong baselines such as Bitter-RF and iBitter-DRLF confirm the value of feature diversity and meta-level learning in this domain.

While peptides with 100\% sequence identity were removed during dataset curation, we further assessed the impact of sequence similarity by applying an 80\% identity threshold. As detailed in Appendix~\ref{appendix:threshold}, this filtering ensured that no test peptide exceeded 80\% identity with any training peptide. Despite a moderate reduction in size and slight imbalance, the model maintained strong performance (95.3\% accuracy, 0.91 MCC), confirming its robustness under stricter evaluation. These results reinforce the importance of similarity-aware benchmarking. Future studies should consider more stringent thresholds (e.g., 70\%) to address residual overlap, especially for short peptides, while balancing the trade-off between data quality and quantity.

Moreover, the limited availability of experimentally validated bitter and non-bitter peptides remains a key challenge for data-driven models. This scarcity motivated our use of an ensemble of lightweight predictors to mitigate overfitting on small datasets. We hope this work emphasizes the need for broader peptide annotation efforts to support future advancements in the field.

This work suggests that ensemble strategies can play a critical role in advancing peptide-based prediction models, particularly when paired with thoughtful feature integration and selection. Future work may extend this modular framework to related tasks such as bitterness intensity prediction, peptide solubility classification, or functional motif detection, potentially incorporating deep end-to-end learning for further automation and scalability.

\section{\textbf{Data Availability}}
\label{sec:dataavailability}

All datasets and source code used in this study are publicly available at our GitHub repository: \href{https://github.com/SarfrazAhmad307/iBitter-Stack}{\texttt{github.com/SarfrazAhmad307/iBitter-Stack}}.  
Additionally, a freely accessible web server for real-time bitter peptide prediction is available at: \href{https://ibitter-stack-webserver.streamlit.app/}{\texttt{ibitter-stack-webserver.streamlit.app}}.

\section*{CRediT authorship contribution statement }

\textbf{Sarfraz Ahmad}: Conceptualization, Investigation, Data curation, Methodology, Software, Writing – original draft, Visualization. \textbf{Momina Ahsan}: Methodology, Software, Visualization, Validation, Writing – original draft, Writing – review \& editing. \textbf{Muhammad Nabeel Asim}: Conceptualization, Project administration, Methodology, Writing – review \& editing. \textbf{Andreas Dengel}: Supervision, Writing – review \& editing. \textbf{Muhammad Imran Malik}: Conceptualization, Project administration, Methodology, Writing – review \& editing, Supervision.

\section*{Funding}

This research did not receive any specific grant from funding agencies in the public, commercial, or not-for-profit sectors.

\bibliographystyle{elsarticle-num} 
\bibliography{cas-refs}

\begin{thebibliography}{10}
\expandafter\ifx\csname url\endcsname\relax
  \def\url#1{\texttt{#1}}\fi
\expandafter\ifx\csname urlprefix\endcsname\relax\def\urlprefix{URL }\fi
\expandafter\ifx\csname href\endcsname\relax
  \def\href#1#2{#2} \def\path#1{#1}\fi

\bibitem{drewnowski2000bitter}
A.~Drewnowski, C.~Gomez-Carneros, Bitter taste, phytonutrients, and the consumer: a review, The American journal of clinical nutrition 72~(6) (2000) 1424--1435.

\bibitem{matoba1972relationship}
T.~Matoba, T.~Hata, Relationship between bitterness of peptides and their chemical structures, Agricultural and Biological Chemistry 36~(8) (1972) 1423--1431.

\bibitem{van2002ftir}
C.~Van Der~Ven, S.~Muresan, H.~Gruppen, D.~B. De~Bont, K.~B. Merck, A.~G. Voragen, Ftir spectra of whey and casein hydrolysates in relation to their functional properties, Journal of Agricultural and Food Chemistry 50~(24) (2002) 6943--6950.

\bibitem{kim2006application}
H.-O. Kim, E.~C. Li-Chan, Application of fourier transform raman spectroscopy for prediction of bitterness of peptides, Applied spectroscopy 60~(11) (2006) 1297--1306.

\bibitem{karametsi2014identification}
K.~Karametsi, S.~Kokkinidou, I.~Ronningen, D.~G. Peterson, Identification of bitter peptides in aged cheddar cheese, Journal of agricultural and food chemistry 62~(32) (2014) 8034--8041.

\bibitem{liu2014identification}
X.~Liu, D.~Jiang, D.~G. Peterson, Identification of bitter peptides in whey protein hydrolysate, Journal of agricultural and food chemistry 62~(25) (2014) 5719--5725.

\bibitem{asim2020k}
M.~N. Asim, M.~I. Malik, A.~Dengel, S.~Ahmed, K-mer neural embedding performance analysis using amino acid codons, in: 2020 International Joint Conference on Neural Networks (IJCNN), IEEE, 2020, pp. 1--8.

\bibitem{asim2022circ}
M.~N. Asim, M.~A. Ibrahim, M.~Imran~Malik, A.~Dengel, S.~Ahmed, Circ-locnet: A computational framework for circular rna sub-cellular localization prediction, International Journal of Molecular Sciences 23~(15) (2022) 8221.

\bibitem{nabeel2023dna}
M.~Nabeel~Asim, M.~Ali~Ibrahim, A.~Fazeel, A.~Dengel, S.~Ahmed, Dna-mp: a generalized dna modifications predictor for multiple species based on powerful sequence encoding method, Briefings in Bioinformatics 24~(1) (2023) bbac546.

\bibitem{asim2025peptide}
M.~N. Asim, T.~Asif, F.~Mehmood, A.~Dengel, Peptide classification landscape: An in-depth systematic literature review on peptide types, databases, datasets, predictors architectures and performance, Computers in Biology and Medicine 188 (2025) 109821.

\bibitem{le2021radiomics}
N.~Q.~K. Le, T.~N.~K. Hung, D.~T. Do, L.~H.~T. Lam, L.~H. Dang, T.-T. Huynh, Radiomics-based machine learning model for efficiently classifying transcriptome subtypes in glioblastoma patients from mri, Computers in Biology and Medicine 132 (2021) 104320.

\bibitem{ramzan2021machine}
Z.~Ramzan, M.~A. Hassan, H.~Asif, A.~Farooq, A machine learning-based self-risk assessment technique for cervical cancer, Current Bioinformatics 16~(2) (2021) 315--332.

\bibitem{kim2006quantitative}
H.-O. Kim, E.~C. Li-Chan, Quantitative structure- activity relationship study of bitter peptides, Journal of agricultural and food chemistry 54~(26) (2006) 10102--10111.

\bibitem{wu2007quantitative}
J.~Wu, R.~E. Aluko, Quantitative structure-activity relationship study of bitter di-and tri-peptides including relationship with angiotensin i-converting enzyme inhibitory activity, Journal of peptide science: an official publication of the European Peptide Society 13~(1) (2007) 63--69.

\bibitem{yin2010studying}
J.~Yin, Y.~Diao, Z.~Wen, Z.~Wang, M.~Li, Studying peptides biological activities based on multidimensional descriptors (e) using support vector regression, International Journal of Peptide Research and Therapeutics 16 (2010) 111--121.

\bibitem{lin2008new}
Z.-h. Lin, H.-x. Long, Z.~Bo, Y.-q. Wang, Y.-z. Wu, New descriptors of amino acids and their application to peptide qsar study, Peptides 29~(10) (2008) 1798--1805.

\bibitem{tong2008novel}
J.~Tong, S.~Liu, P.~Zhou, B.~Wu, Z.~Li, A novel descriptor of amino acids and its application in peptide qsar, Journal of Theoretical Biology 253~(1) (2008) 90--97.

\bibitem{liang2009using}
G.~Liang, L.~Yang, L.~Kang, H.~Mei, Z.~Li, Using multidimensional patterns of amino acid attributes for qsar analysis of peptides, Amino Acids 37 (2009) 583--591.

\bibitem{pripp2007modelling}
A.~H. Pripp, Y.~Ard{\"o}, Modelling relationship between angiotensin-(i)-converting enzyme inhibition and the bitter taste of peptides, Food Chemistry 102~(3) (2007) 880--888.

\bibitem{soltani2013qsbr}
S.~Soltani, H.~Haghaei, A.~Shayanfar, J.~Vallipour, K.~Asadpour~Zeynali, A.~Jouyban, Qsbr study of bitter taste of peptides: application of ga-pls in combination with mlr, svm, and ann approaches, BioMed research international 2013~(1) (2013) 501310.

\bibitem{huang2016bitterx}
W.~Huang, Q.~Shen, X.~Su, M.~Ji, X.~Liu, Y.~Chen, S.~Lu, H.~Zhuang, J.~Zhang, Bitterx: a tool for understanding bitter taste in humans, Scientific Reports 6~(1) (2016) 23450.

\bibitem{dagan2017bitter}
A.~Dagan-Wiener, I.~Nissim, N.~Ben~Abu, G.~Borgonovo, A.~Bassoli, M.~Y. Niv, Bitter or not? bitterpredict, a tool for predicting taste from chemical structure, Scientific Reports 7~(1) (2017) 12074.

\bibitem{charoenkwan2020ibitter}
P.~Charoenkwan, J.~Yana, N.~Schaduangrat, C.~Nantasenamat, M.~M. Hasan, W.~Shoombuatong, ibitter-scm: Identification and characterization of bitter peptides using a scoring card method with propensity scores of dipeptides, Genomics 112~(4) (2020) 2813--2822.

\bibitem{charoenkwan2021bert4bitter}
P.~Charoenkwan, C.~Nantasenamat, M.~M. Hasan, B.~Manavalan, W.~Shoombuatong, Bert4bitter: a bidirectional encoder representations from transformers (bert)-based model for improving the prediction of bitter peptides, Bioinformatics 37~(17) (2021) 2556--2562.

\bibitem{manavalan2019mahtpred}
B.~Manavalan, S.~Basith, T.~H. Shin, L.~Wei, G.~Lee, mahtpred: a sequence-based meta-predictor for improving the prediction of anti-hypertensive peptides using effective feature representation, Bioinformatics 35~(16) (2019) 2757--2765.

\bibitem{manavalan2019meta}
B.~Manavalan, S.~Basith, T.~H. Shin, L.~Wei, G.~Lee, Meta-4mcpred: a sequence-based meta-predictor for accurate dna 4mc site prediction using effective feature representation, Molecular Therapy Nucleic Acids 16 (2019) 733--744.

\bibitem{charoenkwan2013hcs}
P.~Charoenkwan, E.~Hwang, R.~W. Cutler, H.-C. Lee, L.-W. Ko, H.-L. Huang, S.-Y. Ho, Hcs-neurons: identifying phenotypic changes in multi-neuron images upon drug treatments of high-content screening, BMC bioinformatics 14 (2013) 1--15.

\bibitem{charoenkwan2019iqsp}
P.~Charoenkwan, N.~Schaduangrat, C.~Nantasenamat, T.~Piacham, W.~Shoombuatong, iqsp: a sequence-based tool for the prediction and analysis of quorum sensing peptides via chou’s 5-steps rule and informative physicochemical properties, International Journal of Molecular Sciences 21~(1) (2019) 75.

\bibitem{liu2020imrm}
K.~Liu, W.~Chen, imrm: a platform for simultaneously identifying multiple kinds of rna modifications, Bioinformatics 36~(11) (2020) 3336--3342.

\bibitem{liu2020irna5hmc}
Y.~Liu, D.~Chen, R.~Su, W.~Chen, L.~Wei, irna5hmc: the first predictor to identify rna 5-hydroxymethylcytosine modifications using machine learning, Frontiers in bioengineering and biotechnology 8 (2020) 227.

\bibitem{charoenkwan2021ibitter}
P.~Charoenkwan, C.~Nantasenamat, M.~M. Hasan, M.~A. Moni, P.~Lio’, W.~Shoombuatong, ibitter-fuse: a novel sequence-based bitter peptide predictor by fusing multi-view features, International Journal of Molecular Sciences 22~(16) (2021) 8958.

\bibitem{jiang2022identify}
J.~Jiang, X.~Lin, Y.~Jiang, L.~Jiang, Z.~Lv, Identify bitter peptides by using deep representation learning features, International journal of molecular sciences 23~(14) (2022) 7877.

\bibitem{du2023unidl4biopep}
Z.~Du, X.~Ding, Y.~Xu, Y.~Li, Unidl4biopep: a universal deep learning architecture for binary classification in peptide bioactivity, Briefings in Bioinformatics 24~(3) (2023) bbad135.

\bibitem{zhang2023bitter}
Y.-F. Zhang, Y.-H. Wang, Z.-F. Gu, X.-R. Pan, J.~Li, H.~Ding, Y.~Zhang, K.-J. Deng, Bitter-rf: a random forest machine model for recognizing bitter peptides, Frontiers in medicine 10 (2023) 1052923.

\bibitem{lv2025ibitter}
J.~Lv, A.~Geng, Z.~Pan, L.~Wei, Q.~Zou, Z.~Zhang, F.~Cui, ibitter-gre: A novel stacked bitter peptide predictor with esm-2 and multi-view features, Journal of Molecular Biology (2025) 169005.

\bibitem{xu2019quantitative}
B.~Xu, H.~Y. Chung, Quantitative structure--activity relationship study of bitter di-, tri-and tetrapeptides using integrated descriptors, Molecules 24~(15) (2019) 2846.

\bibitem{vasylenko2015scmpsp}
T.~Vasylenko, Y.-F. Liou, H.-A. Chen, P.~Charoenkwan, H.-L. Huang, S.-Y. Ho, Scmpsp: Prediction and characterization of photosynthetic proteins based on a scoring card method, in: BMC bioinformatics, Vol.~16, Springer, 2015, pp. 1--16.

\bibitem{vasylenko2016scmbyk}
T.~Vasylenko, Y.-F. Liou, P.-C. Chiou, H.-W. Chu, Y.-S. Lai, Y.-L. Chou, H.-L. Huang, S.-Y. Ho, Scmbyk: prediction and characterization of bacterial tyrosine-kinases based on propensity scores of dipeptides, BMC bioinformatics 17 (2016) 203--217.

\bibitem{charoenkwan2013scmcrys}
P.~Charoenkwan, W.~Shoombuatong, H.-C. Lee, J.~Chaijaruwanich, H.-L. Huang, S.-Y. Ho, Scmcrys: predicting protein crystallization using an ensemble scoring card method with estimating propensity scores of p-collocated amino acid pairs, PloS one 8~(9) (2013) e72368.

\bibitem{liou2015scmmtp}
Y.-F. Liou, T.~Vasylenko, C.-L. Yeh, W.-C. Lin, S.-H. Chiu, P.~Charoenkwan, L.-S. Shu, S.-Y. Ho, H.-L. Huang, Scmmtp: identifying and characterizing membrane transport proteins using propensity scores of dipeptides, BMC genomics 16 (2015) 1--14.

\bibitem{zheng2018bitter}
S.~Zheng, M.~Jiang, C.~Zhao, R.~Zhu, Z.~Hu, Y.~Xu, F.~Lin, e-bitter: bitterant prediction by the consensus voting from the machine-learning methods, Frontiers in chemistry 6 (2018) 82.

\bibitem{minkiewicz2008biopep}
P.~Minkiewicz, J.~Dziuba, A.~Iwaniak, M.~Dziuba, M.~Darewicz, Biopep database and other programs for processing bioactive peptide sequences, Journal of AOAC International 91~(4) (2008) 965--980.

\bibitem{gautam2013silico}
A.~Gautam, K.~Chaudhary, R.~Kumar, A.~Sharma, P.~Kapoor, A.~Tyagi, O.~S. D. D. C.~I. osdd. net, G.~P. Raghava, In silico approaches for designing highly effective cell penetrating peptides, Journal of translational medicine 11 (2013) 1--12.

\bibitem{kumar2015silico}
R.~Kumar, K.~Chaudhary, J.~Singh~Chauhan, G.~Nagpal, R.~Kumar, M.~Sharma, G.~P. Raghava, An in silico platform for predicting, screening and designing of antihypertensive peptides, Scientific reports 5~(1) (2015) 12512.

\bibitem{tyagi2013silico}
A.~Tyagi, P.~Kapoor, R.~Kumar, K.~Chaudhary, A.~Gautam, G.~Raghava, In silico models for designing and discovering novel anticancer peptides, Scientific reports 3~(1) (2013) 2984.

\bibitem{charoenkwan2020idppiv}
P.~Charoenkwan, S.~Kanthawong, C.~Nantasenamat, M.~M. Hasan, W.~Shoombuatong, idppiv-scm: a sequence-based predictor for identifying and analyzing dipeptidyl peptidase iv (dpp-iv) inhibitory peptides using a scoring card method, Journal of proteome research 19~(10) (2020) 4125--4136.

\bibitem{charoenkwan2020iumami}
P.~Charoenkwan, J.~Yana, C.~Nantasenamat, M.~M. Hasan, W.~Shoombuatong, iumami-scm: a novel sequence-based predictor for prediction and analysis of umami peptides using a scoring card method with propensity scores of dipeptides, Journal of Chemical Information and Modeling 60~(12) (2020) 6666--6678.

\bibitem{wei2021computational}
L.~Wei, W.~He, A.~Malik, R.~Su, L.~Cui, B.~Manavalan, Computational prediction and interpretation of cell-specific replication origin sites from multiple eukaryotes by exploiting stacking framework, Briefings in Bioinformatics 22~(4) (2021) bbaa275.

\bibitem{chung2020characterization}
C.-R. Chung, T.-R. Kuo, L.-C. Wu, T.-Y. Lee, J.-T. Horng, Characterization and identification of antimicrobial peptides with different functional activities, Briefings in bioinformatics 21~(3) (2020) 1098--1114.

\bibitem{chaudhary2016web}
K.~Chaudhary, R.~Kumar, S.~Singh, A.~Tuknait, A.~Gautam, D.~Mathur, P.~Anand, G.~C. Varshney, G.~P. Raghava, A web server and mobile app for computing hemolytic potency of peptides, Scientific reports 6~(1) (2016) 22843.

\bibitem{lin2023evolutionary}
Z.~Lin, H.~Akin, R.~Rao, B.~Hie, Z.~Zhu, W.~Lu, N.~Smetanin, R.~Verkuil, O.~Kabeli, Y.~Shmueli, et~al., Evolutionary-scale prediction of atomic-level protein structure with a language model, Science 379~(6637) (2023) 1123--1130.

\bibitem{mcinnes2018umap}
L.~McInnes, J.~Healy, J.~Melville, Umap: Uniform manifold approximation and projection for dimension reduction, arXiv preprint arXiv:1802.03426 (2018).

\bibitem{hinton2008visualizing}
G.~Hinton, L.~Van Der~Maaten, Visualizing data using t-sne journal of machine learning research, Journal of Machine Learning Research 9~(2579-2605) (2008) 4.

\bibitem{dvzeroski2004combining}
S.~D{\v{z}}eroski, B.~{\v{Z}}enko, Is combining classifiers with stacking better than selecting the best one?, Machine learning 54 (2004) 255--273.

\end{thebibliography}

\appendix

\begin{figure*}[!t]
    \centering
    \includegraphics[width=0.7\textwidth]{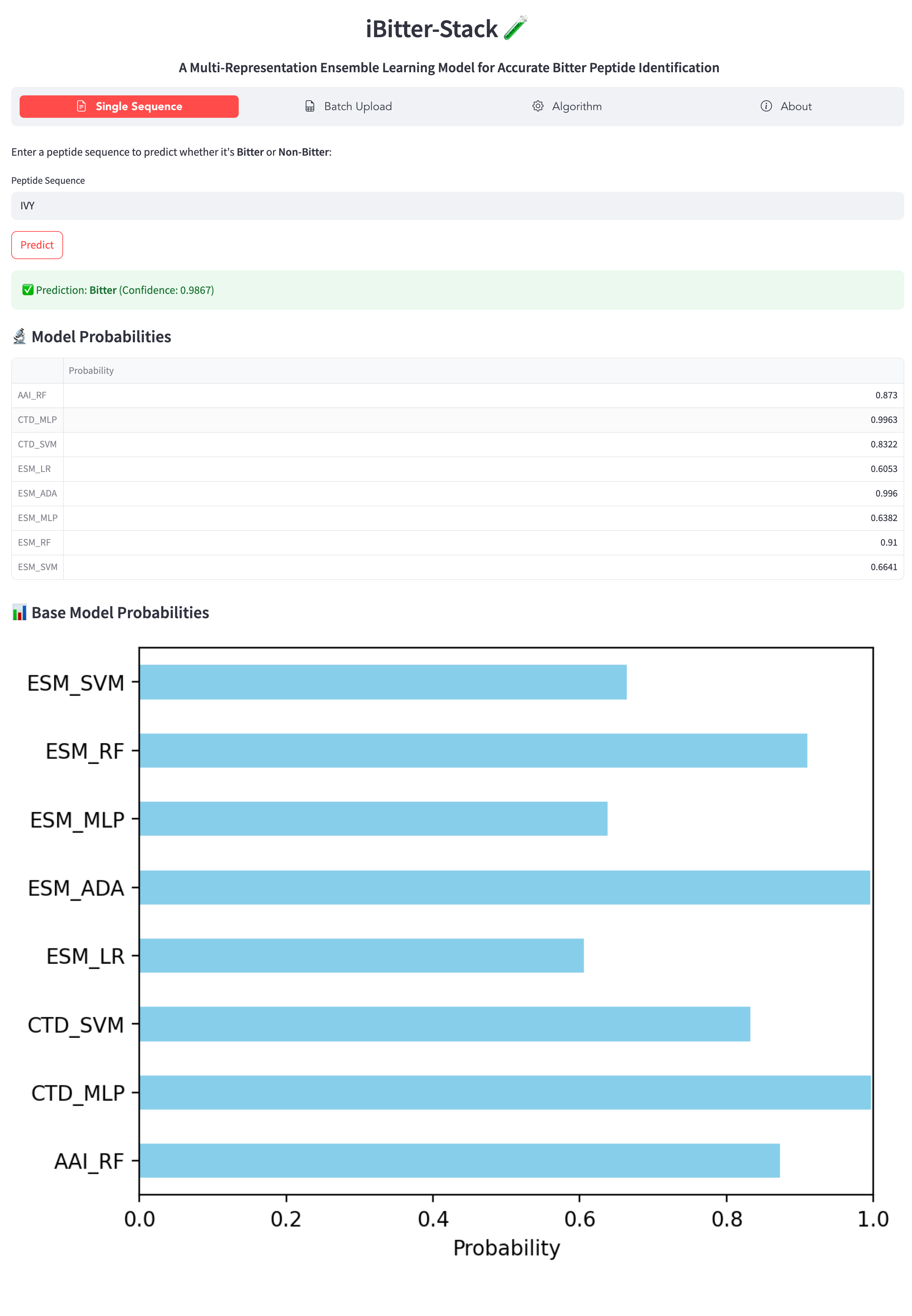} 
    \caption{Single sequence input mode on iBitter-Stack along with model probabilities contributed by the eight base learners.}
    \label{fig:singleseq-result}
\end{figure*}

\begin{figure*}[!t]
    \centering
    \includegraphics[width=0.8\textwidth]{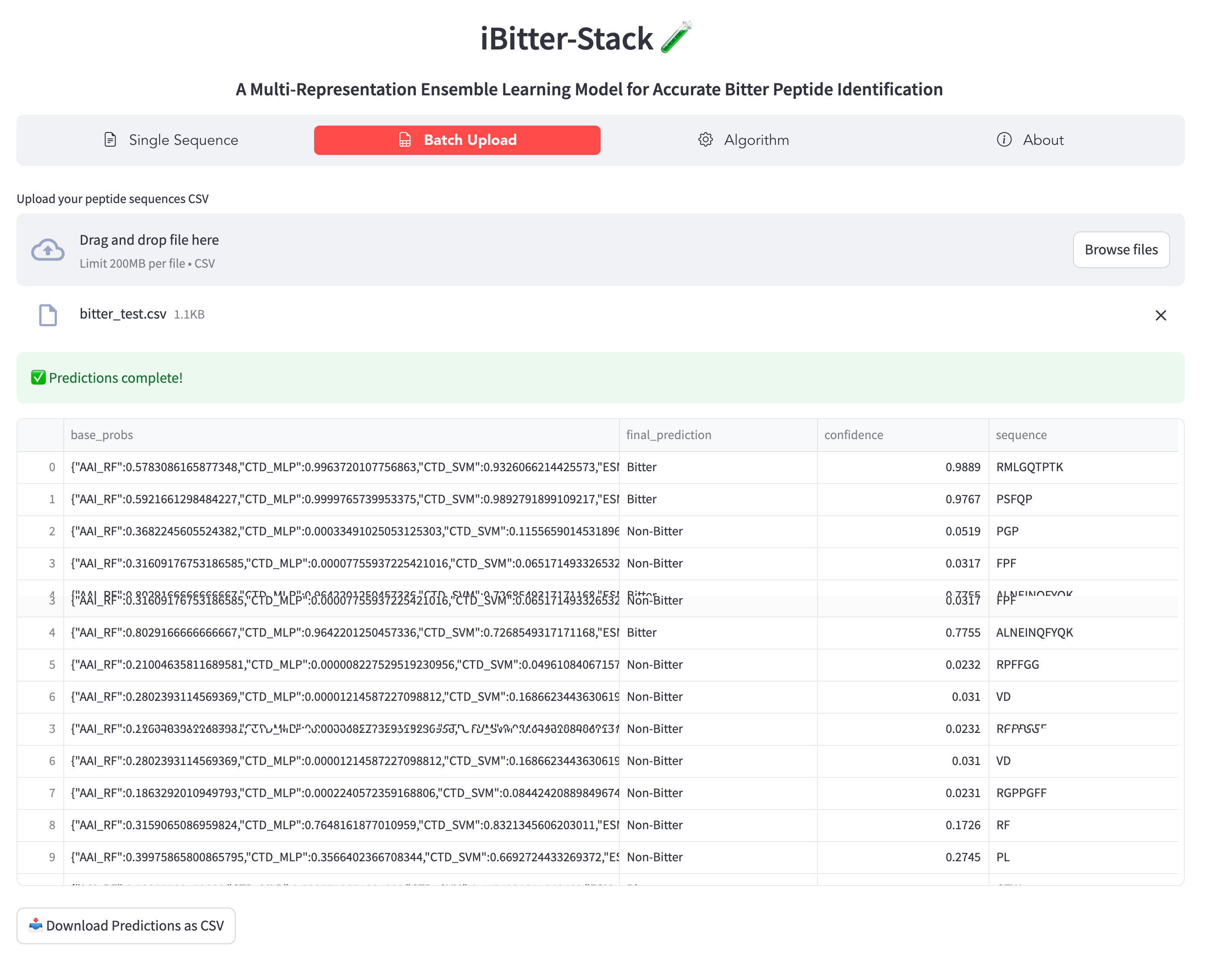} 
    \caption{Batch Upload interface showing successful CSV file submission along with partial table view of batch-wise predictions including base learner probabilities.}
    \label{fig:batch-upload}
\end{figure*}

\section{Additional Experiment: Sequence Similarity Filtering}
\label{appendix:threshold}

To further address the concern regarding sequence redundancy between training and testing sets, we conducted a follow-up experiment in which sequence similarity was explicitly controlled. While peptides with 100\% identity were previously removed, more subtle similarities may still bias performance metrics, especially for short peptides. Consequently, we employed a more stringent similarity reduction strategy using an 80\% identity threshold.

\subsection{Rationale for Similarity Threshold Selection}

The selection of the 80\% similarity cutoff was empirically guided. Initial trials with a 90\% threshold led to the removal of only a marginal number of peptides, thus offering limited improvement in redundancy reduction. In contrast, more aggressive thresholds such as 70\%, 60\%, and 50\% resulted in a drastic reduction of data, which is particularly problematic given the limited availability of experimentally validated bitter and nonbitter peptides. For example, filtering at 60\% retained fewer than 200 bitter sequences in training, thereby undermining statistical robustness. The 80\% threshold therefore strikes a practical balance between eliminating highly similar sequences and preserving sufficient dataset size for reliable model development.

\subsection{Similarity Filtering Procedure}

Sequence similarity filtering was performed using a custom Python pipeline based on global pairwise alignment via the \texttt{Biopython}\footnote{https://biopython.org/wiki/Packages} package. The filtering process involved three main stages:

\begin{enumerate}
    \item \textbf{Within set filtering:} Peptides in each of the training and testing sets were compared pairwise. Any sequence found to share at least 80\% identity with another sequence in the same set was removed.

    \item \textbf{Across set filtering:} All sequences in the filtered test set were compared against the filtered training set. Any test sequence sharing at least 80\% identity with a training sequence was discarded to ensure complete disjointedness.

    \item \textbf{Export:} The resulting datasets were saved for training and evaluation.
\end{enumerate}

\noindent
The impact of this filtering is summarized below:
\begin{itemize}
    \item \textbf{Original dataset size:} Training set = 512 peptides (256 bitter, 256 nonbitter), Test set = 128 peptides (64 bitter, 64 nonbitter)
    \item \textbf{Filtered dataset size:} Training set = 428 peptides (219 bitter, 209 nonbitter), Test set = 86 peptides (44 bitter, 42 nonbitter)
\end{itemize}

This procedure resulted in a moderate reduction in overall dataset size (from 640 to 514 peptides), along with a minor imbalance in the class distribution. Nevertheless, both sets retained sufficient representation of each class to support meaningful evaluation.

\subsection{Updated Meta Learner Results}

The meta learning framework was retrained on the filtered dataset. The top eight base learners selected for the new ensemble included two models not present in the original configuration: \texttt{BPNC\_RF} and \texttt{ESM\_KNN}. This change highlights the sensitivity of base learner selection to the underlying distribution of sequence patterns.

\begin{table}[!ht]
\caption{Performance of Proposed Model Before and After Sequence Similarity Filtering (Independent Test Set)}
\label{tab:filtered-comparison}
\centering
\resizebox{\columnwidth}{!}{
\begin{tabular}{lccccc}
\toprule
\textbf{Model} & \textbf{Acc (\%)} & \textbf{Sn (\%)} & \textbf{Sp (\%)} & \textbf{MCC} & \textbf{AUROC} \\
\midrule
Proposed (Unfiltered) & \textbf{96.1} & 95.4 & \textbf{97.2} & \textbf{0.92} & \textbf{0.98} \\
Proposed (Filtered, 80\%) & 95.3 & 95.3 & 95.3 & 0.91 & 0.98 \\
\bottomrule
\end{tabular}}
\end{table}

The comparison in Table~\ref{tab:filtered-comparison} highlights the stability of the proposed ensemble framework under stricter sequence similarity constraints. After filtering with an 80\% identity threshold, the model retained strong performance, achieving an accuracy of 95.3\%, an MCC of 0.91, and an AUROC of 0.98. These results are only marginally lower than the unfiltered case (96.1\% accuracy and 0.92 MCC), confirming that the predictive capacity of the framework does not rely on redundancy between training and testing sets. Importantly, the more balanced sensitivity and specificity observed in the filtered setting suggests that the ensemble maintains its generalization ability even when the dataset becomes smaller and slightly imbalanced. This finding demonstrates methodological robustness and provides greater confidence that the model’s high performance reflects genuine learning of discriminative sequence patterns rather than potential overlap effects. Moreover, this refined evaluation protocol improves methodological rigor and enables more reliable comparisons in future benchmarking studies.

\section{Web Server Interface}
\label{appendix:webserver}

The iBitter-Stack web server provides an intuitive user interface for predicting the bitterness of peptide sequences. In the Single Sequence mode, users can input an individual peptide sequence and receive a prediction result along with the model’s confidence. Figure~\ref{fig:singleseq-result} shows the input interface where the peptide sequence "IVY" is entered. The prediction output indicates the peptide is predicted as "Bitter" with 98.67\% confidence. Additionally, Fig.~\ref{fig:singleseq-result} also displays the confidence scores assigned by each of the eight base learners used in the ensemble, aiding interpretability.

The Batch Upload mode allows users to submit a CSV file containing multiple peptide sequences for simultaneous prediction. This feature is particularly useful for high-throughput screening scenarios. As illustrated in Fig.~\ref{fig:batch-upload}, users can drag and drop a CSV file (e.g., \texttt{bitter\_test.csv}) to begin the prediction process. Once completed, results including base learner probabilities, final prediction labels, and confidence scores are displayed in a downloadable table. The web server is publicly available at
\url{https://ibitter-stack-webserver.streamlit.app/}.

\end{document}